\documentclass[a4paper, 11pt]{article}

\usepackage{fullpage}
\usepackage{amsmath,amssymb,float,graphicx,multirow,enumerate}
\usepackage{hyperref}
\hypersetup{colorlinks = true, linktocpage = true, bookmarksopen = true, linkcolor = blue, urlcolor=blue, citecolor = blue, allcolors=blue}
\usepackage[caption=false]{subfig}
\usepackage[font=footnotesize]{caption}
\usepackage[numbers,sort&compress]{natbib}

\newcommand{\revision}[1]{#1}

\begin{document}

\title{Characteristic time scales for diffusion processes through layers and across interfaces}

\author{Elliot\hspace{0.1cm}J.\hspace{0.1cm}Carr\footnote{\href{mailto:elliot.carr@qut.edu.au}{elliot.carr@qut.edu.au}}\\
\small School of Mathematical Sciences, Queensland University of Technology (QUT), Brisbane, Australia.}
\date{}

\maketitle

\begin{abstract}
This paper presents a simple tool for characterising the timescale for continuum diffusion processes through layered heterogeneous media. This mathematical problem is motivated by several practical applications such as heat transport in composite materials, flow in layered aquifers and drug diffusion through the layers of the skin. In such processes, the physical properties of the medium vary across layers and internal boundary conditions apply at the interfaces between adjacent layers. To characterise the timescale, we use the concept of mean action time, which provides the mean timescale at each position in the medium by utilising the fact that the transition of the transient solution of the underlying partial differential equation model, from initial state to steady state, can be represented as a cumulative distribution function of time. Using this concept, we define the characteristic timescale for a multilayer diffusion process as the maximum value of the mean action time across the layered medium. For given initial conditions and internal and external boundary conditions, this approach leads to simple algebraic expressions for characterising the timescale that depend on the physical and geometrical properties of the medium, such as the diffusivities and lengths of the layers. Numerical examples demonstrate that these expressions provide useful insight into explaining how the parameters in the model affect the time it takes for a multilayer diffusion process to reach steady state.
\end{abstract}




\section{Introduction}
When a physical system governed by diffusive transport is disturbed from an initial state of equilibrium by a sudden change in boundary conditions, it takes an infinite amount of time for the physical system to reach a new equilibrium state \cite{mcnabb_1991,landman_2000}. While this is strictly true (as the transient solution of the underlying partial differential equation model approaches the steady state solution exponentially), there exists a finite time at which the diffusive process is effectively at steady state (i.e. the difference between the transient and steady state solutions is less than a small specified tolerance) \cite{carr_2017b}. In a one-dimensional medium, these so-called finite transition times are often characterised by a timescale proportional to $\ell^{2}/D$, where $D$ is the diffusivity and $\ell$ is the length of the medium \cite{crank_1975,hickson_2009_part1,carr_2017b}. Such an expression provides a simple formula for the timescale of the diffusive process and its form is intuitive since we expect diffusive processes to take longer for smaller $D$ (fixed $\ell$) and larger $\ell$ (fixed $D$). However, two obvious limitations of this result are that it applies only in the case of a homogeneous medium, where the diffusivity is spatially constant, and it is only valid for certain choices of boundary conditions \cite{crank_1975}.

The focus of this paper is characterising the timescale of diffusion processes through layered heterogeneous media, where the physical properties of the medium such as the diffusivity vary across the layers. This mathematical problem is motivated by several practical applications. In drug delivery systems involving drug release from multilayer spherical capsules \cite{kaoui_2018} or drug diffusion through the layers of the skin \cite{pontrelli_2018}, a characteristic timescale allows one to assess the key parameters influencing the release performance of the delivery system. In hydrology applications involving flow in layered aquifers \cite{liu_1998}, characterisation of the response time is useful for determining when a simpler steady state model can be used \cite{simpson_2013,carr_2018a}. Finally, a characteristic timescale for multilayer heat conduction enables the thermal response of composite materials to be assessed \cite{absi_2005}.

For multilayer diffusion processes, simple characterisation of the timescale in terms of the parameters in the model is not straightforward. This is evident when considering a diffusion process in a medium consisting of two layers of lengths $\ell_{1}$ and $\ell_{2}$ and diffusivities $D_{1}$ and $D_{2}$. Present in this diffusion process are two obvious timescales $\ell_{1}^{2}/D_{1}$ and $\ell_{2}^{2}/D_{2}$, however, it is not immediately obvious how these timescales should be combined into a single timescale representative of the entire physical system. The aim of this paper is thus to develop simple formulas for characterising multilayer diffusion processes analogous to the expression $\ell^{2}/D$ for single-layer diffusion.

Two common approaches for characterising the timescale of diffusion processes are the concepts of \textit{time lag} \cite{crank_1975,frisch_1957, ash_1965,rutherford_1997,hickson_thesis_2010} and \textit{effective time constant} \cite{collins_1980,simon_2009,pontrelli_2018}. The first approach is based on the long-time asymptotic behaviour of the total amount of diffusing substance that has passed through the medium by a given time. This asymptotic behaviour takes the form of a linear function of time with the intercept of this line with the time axis referred to as the time lag \cite{crank_1975}. Although the time lag can be calculated without requiring the transient solution of the underlying partial differential equation model \cite{frisch_1957}, it produces a characteristic timescale for multilayer diffusion processes \cite{ash_1965} that does not account for the fact that permuting the layer ordering affects the time required to reach steady state \cite{hickson_thesis_2010}. The effective time constant, on the other hand, is defined as the mean time calculated under the assumption that time is distributed according to a specific probability density function representing the transition of the diffusive process from initial to steady state \cite{collins_1980,simon_2009}. Evaluating the mean is then typically carried out in Laplace transform space, which requires solving for the Laplace transform of the transient solution of the underlying partial differential equation \cite{simon_2016}. While this approach is straightforward for single-layer (homogeneous) diffusion and explains how the timescale depends on position \cite{collins_1980}, deriving the Laplace transform of the transient solution in each layer becomes tedious as the number of layers is increased and results in complicated expressions relating the effective time constant to the parameters in the model \cite{pontrelli_2018}.

In this paper, we use the concept of \textit{mean action time} (MAT) \cite{mcnabb_1991,landman_2000,ellery_2013b,simpson_2013,carr_2017b} to characterise the timescale for multilayer diffusive processes. The attraction of working with MAT is that it combines two desirable properties of the time lag and effective time constant: (i) MAT can be calculated without the transient solution of the underlying partial differential equation model (property of time lag) (ii) MAT explains how the timescale varies with position (property of effective time constant). Similarly to the effective time constant, MAT is defined as a mean time calculated under the assumption that time follows a specific probability distribution \cite{ellery_2012a,simpson_2013,carr_2017b}, however, the corresponding probability density function lends itself to simpler analysis. Within the MAT framework, calculating the higher-order moments of the distribution also follows in a straightforward manner \cite{carr_2017b} and these can be used to calculate the time required for the transient solution to transition to within a small specified tolerance of the steady-state solution \cite{carr_2017b,carr_2018a}.

The application of MAT concepts to multilayer diffusion processes has previously been carried out by \citet{hickson_thesis_2010}. However, in that work approximate numerical values are given for specific parameter values only, calculated by first solving for the transient solution of the multilayer diffusion model. A key contribution of this paper is that we show how MAT can be calculated exactly for layered diffusion problems without requiring the transient solution of the underlying partial differential equation model. Ultimately, our analysis leads to simple algebraic expressions for characterising the timescale that depend on the physical and geometrical parameters in the model. As we demonstrate, these results provide a straightforward way to characterise and compare the timescales of different multilayer diffusion processes. 

The remaining sections of this paper are organised in the following way. In the next section, we explain the concept of MAT in more detail, include a brief comparison to the concept of effective time constant and provide a geometrical interpretation of MAT that explains why it is useful for characterising the time required to reach steady state. Section \ref{sec:multilayer} presents the multilayer diffusion model considered in this paper and develops the procedure for calculating MAT within each layer without computing the transient solution. In Section \ref{sec:results}, the new approach is applied to some general multilayer diffusion problems to derive simple algebraic expressions for characterising the timescale. Some numerical examples are then presented that illustrate how these simple expressions provide useful insight into how the parameters in the model affect how long it takes for a multilayer diffusion process to reach steady state. Finally, in Section \ref{sec:conclusions}, we summarise the main contributions of the work and discuss extensions of the analysis.

\begin{figure*}[!t]
\centering
\includegraphics[width=\textwidth]{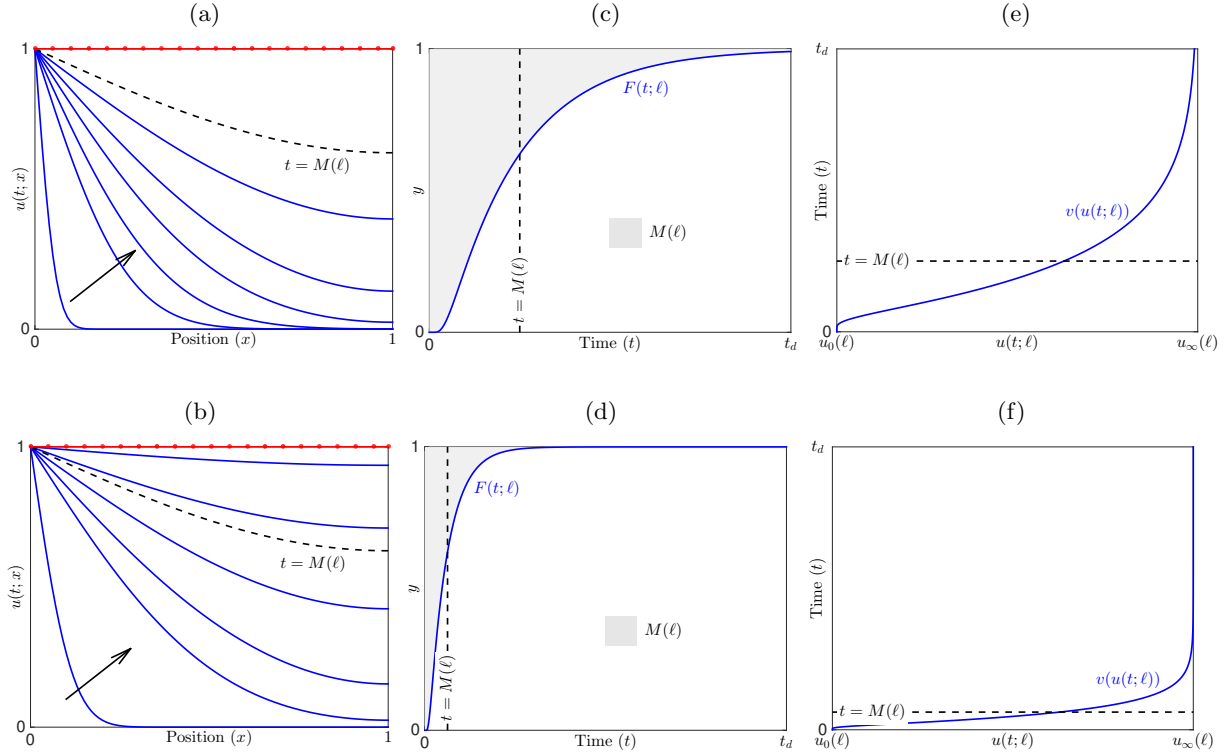}
\caption{\textbf{Geometrical interpretations of MAT.}~ Example curves for the diffusion model described by Eqs (\ref{eq:pde}) and (\ref{eq:ic_bcs}) with $u_{b} = 1$ and $\ell = 1$ \revision{for $D = 0.1$ [(a),(c),(e)] and  $D = 0.4$ [(b),(d),(f)]}: (a)--(b) Plot of the transient solution $u(t;x)$ at $t = [0.01,0.2,0.4,0.8,1.5,3.0]$ \revision{(blue solid lines)} and at the characteristic timescale $t = M(\ell) = \ell^{2}/(2D)$ as in Eq.~(\ref{eq:characteristic_example}) (black dashed line). Each plot also includes the steady state solution $u_{\infty}(x) = 1$ \revision{(red dash-dot line)} and a black arrow indicating the direction of increasing time $t$. (c)--(d) Plot of the cumulative distribution function $F(t;x)$ (\ref{eq:cdf}) at $x = \ell$ and $0\leq t \leq t_{d} = 20$. The shaded area depicts the geometrical interpretation of MAT described in Eq.~(\ref{eq:mat3}) while the vertical black line represents MAT at \revision{$x=\ell$. (e)--(f)} Representation of MAT as the mean value of the inverse function $v(u(t;x))$ that maps the transient solution $u(t;x)$ to $t$ as described in Eq.~(\ref{eq:mat4}). The inverse function $v(u(t;x))$ is plotted at $x = \ell$ and MAT at $x = \ell$ is represented by a horizontal dashed black line.}
\label{fig:mat_representations}
\end{figure*}

\section{Mean action time}
\label{sec:mean_action_time}
To illustrate the concept of MAT, consider the linear diffusion equation
\begin{gather}
\label{eq:pde}
\frac{\partial u}{\partial t} = D\frac{\partial^{2}u}{\partial x^{2}},\quad 0 < x < \ell,\quad t > 0,
\end{gather}
where $u(t;x)$ is the transient solution at position $x$ and time $t$ and $\ell$ is the length of the medium (Fig.~\ref{fig:mat_representations}ab). Note that we use the notation $u(t;x)$ rather than $u(x,t)$ to emphasise that $x$ is treated as a parameter. Let $u_{0}(x)$ be the initial solution and $u_{\infty}(x)$ be the steady state solution and define \cite{ellery_2012a,simpson_2013,carr_2017b}:
\begin{gather}
\label{eq:cdf}
F(t; x) := 1 - \frac{u(t;x) - u_{\infty}(x)}{u_{0}(x) - u_{\infty}(x)},
\end{gather}
which satisfies $F(0; x) = 0$ and $\lim_{t\rightarrow\infty} F(t;x) = 1$. Provided the transition of $u(t;x)$ from $t = 0$ to $t\rightarrow\infty$ is monotone, then $F(t;x)$ defines a cumulative distribution function of time $t$, parameterised in terms of $x$ (Fig.~\ref{fig:mat_representations}cd) \cite{ellery_2012a,simpson_2013,carr_2017b}. Within this framework, MAT at position $x$, $M(x)$, is defined as the mean or expected time \cite{ellery_2012a,simpson_2013,carr_2017b}:
\begin{gather}
\label{eq:mat}
M(x) = \int_{0}^{\infty} tf(t;x)\,\textrm{d}t,
\end{gather}
where $f(t;x)$ is the corresponding probability density function:
\begin{align}
\nonumber
f(t;x) &= \frac{\partial F(t;x)}{\partial t}\\\label{eq:pdf} &= \frac{1}{u_{\infty}(x)-u_{0}(x)}\frac{\partial}{\partial t}\left[u(t;x) - u_{\infty}(x)\right].
\end{align}
Substituting (\ref{eq:pdf}) into (\ref{eq:mat}), applying integration by parts and noting that $\lim_{t\rightarrow\infty} t\left[u(t;x)-u_{\infty}(x)\right] = 0$ yields \cite{landman_2000,ellery_2012b,carr_2017c}:
\begin{gather}
\label{eq:mat2}
M(x) = \int_{0}^{\infty} \frac{u_{\infty}(x)-u(t;x)}{u_{\infty}(x)-u_{0}(x)}\,\textrm{d}t.
\end{gather}
In this paper, we define the characteristic timescale of a diffusion process as the maximum value of MAT:
\begin{align}
\label{eq:characteristic_timescale}
\tau = \max_{x\in[0,\ell]}M(x).
\end{align}
Our justification for this choice is provided by two geometrical interpretations of MAT. Firstly, combining (\ref{eq:cdf}) and (\ref{eq:mat2}) yields:
\begin{gather}
\label{eq:mat3}
M(x) = \int_{0}^{\infty} 1 - F(t;x)\,\textrm{d}t,
\end{gather}
and hence $M(x)$ is precisely the area between the curves $y = F(t; x)$ and $y = 1$ as depicted in Figs.~\ref{fig:mat_representations}cd (if $t_{d}\rightarrow\infty$). In statistics, this is a well known result for relating the mean and cumulative distribution function of a continuous random variable with non-negative support \cite[pg.~84]{ibe_2014}. With this interpretation, it is evident that MAT provides a useful measure of the time required for a diffusive process to reach steady state because the area between $y = F(t;x)$ and $y = 1$ will be larger for slower transitions (Fig.~\ref{fig:mat_representations}c) and smaller for faster transitions (Fig.~\ref{fig:mat_representations}d). Therefore, in addition to providing the characteristic timescale of the diffusive process, Eq.~(\ref{eq:characteristic_timescale}) also provides a useful indicator for how long it takes for the diffusive process to reach steady state. Another geometrical interpretation of MAT justifying (\ref{eq:characteristic_timescale}) stems from noting that Eq.~(\ref{eq:mat2}) is equivalent to:
\begin{gather}
\label{eq:mat4}
M(x) = \frac{1}{u_{\infty}(x)-u_{0}(x)}\int_{u_{0}(x)}^{u_{\infty}(x)} v(u(t;x))\,\textrm{d}u.
\end{gather}
Here, we see that MAT is equal to the mean value of the inverse function $v$ that maps the transient solution $u(t;x)$ to $t$ for a given $x$, that is $v(u(t;x)) = t$ for all $t>0$ (Fig.~\ref{fig:mat_representations}ef).

As previously mentioned, a closely related idea to MAT is the concept of an \textit{effective time constant}. As mentioned earlier, both approaches involve calculating a mean time with the difference lying in the construction of the governing probability distribution from the transient solution $u(t;x)$. Specifically, the effective time constant, $t_{\mathrm{eff}}(x)$, is defined as \cite{collins_1980,simon_2009,pontrelli_2018}:
\begin{gather}
\label{eq:etc}
t_{\mathrm{eff}}(x) = \int_{0}^{\infty} t\widetilde{f}(t;x)\,\textrm{d}t,
\end{gather}
with probability density function:
\begin{gather}
\label{eq:pdf_etc}
\widetilde{f}(t;x) := \frac{u_{\infty}(x)-u(t;x)}{\int_{0}^{\infty} u_{\infty}(x)-u(t;x)\,\mathrm{d}t}.
\end{gather}
Calculating (\ref{eq:etc}) is typically carried out using $\overline{U}(s; x)$, the Laplace transform of the transient solution $u(t;x)$ \cite{collins_1980,simon_2009}. This is achieved via the following alternative representation of Eqs.~(\ref{eq:etc})--(\ref{eq:pdf_etc}) derived using properties of the Laplace transform (see \cite{simon_2009}):
\begin{gather}
\label{eq:etc2}
t_{\mathrm{eff}}(x) = \frac{\lim\limits_{s\rightarrow 0}\left[\dfrac{u_{\infty}(x)}{s^{2}} + \dfrac{\mathrm{d}\overline{U}}{\mathrm{d}s}\right]}{\lim\limits_{s\rightarrow 0}\left[\dfrac{u_{\infty}(x)}{s} - \overline{U}(s; x)\right]}.
\end{gather}
Evaluating (\ref{eq:etc2}) is then typically performed using the series expansion of $\overline{U}(s; x)$ about $s = 0$ \cite{collins_1980,simon_2009}. This is because if $\overline{U}(s;x)$ \revision{permits} a series expansion of the form $\overline{U}(s; x) = u_{\infty}(x)s^{-1} + b(x) + c(x)s + O(s^{2})$, then evaluating (\ref{eq:etc2}) reduces to computing $t_{\mathrm{eff}}(x) = -c(x)/b(x)$ \cite{collins_1980,simon_2016}.
 
As we will demonstrate, an advantage of working with MAT (\ref{eq:mat2}) over the effective time constant (\ref{eq:etc2}) is that it can be calculated without computing the transient solution $u(t;x)$ or its Laplace transform $\overline{U}(s; x)$ \cite{landman_2000,ellery_2012a,simpson_2013,carr_2017b}. This becomes increasingly important for multilayer problems as solving for the Laplace transform solution is time-consuming and leads to complicated expressions for $t_{\mathrm{eff}}(x)$ \cite{pontrelli_2018}. Conversely, as we will see in the next section, simple algebraic expressions arise when using MAT.
This reason for this is that the function $\overline{M}(x) = M(x)h(x)$, where $h(x) = u_{\infty}(x) - u_{0}(x)$, satisfies the differential equation \cite{simpson_2013,carr_2017b}:
\begin{gather}
\label{eq:mat_de}
D\overline{M}''(x) = u_{0}(x)-u_{\infty}(x),
\end{gather}
subject to appropriately-defined boundary conditions derived by considering the boundary conditions satisfied by the transient and steady state solutions, $u(t;x)$ and $u_{\infty}(x)$, and making use of the definition (\ref{eq:mat2}) \cite{simpson_2013,carr_2017b}.

For example, for the following initial and boundary conditions:
\begin{gather}
\label{eq:ic_bcs}
u(x,0) = 0,\quad u(0,t) = u_{b},\quad \frac{\partial u}{\partial x}(\ell,t) = 0,
\end{gather}
the differential equation (\ref{eq:mat_de}) is paired with the following derived boundary conditions: $\overline{M}(0) = 0$ and $\overline{M}'(\ell) = 0$ \cite{carr_2017b}. In this case, with $u_{0}(x) = 0$ and $u_{\infty}(x) = u_{b}$, MAT takes the form of 
\begin{gather*}
M(x) = \frac{x(2\ell-x)}{2D}.
\end{gather*} 
This function increases monotonically with increasing $x\in[0,\ell]$, which makes intuitive sense as larger values of $x$ take longer to be affected by the disturbance introduced via the boundary condition at $x = 0$. Maximising $M(x)$ over $[0,\ell]$ produces a characteristic timescale (\ref{eq:characteristic_timescale}) of
\begin{gather}
\label{eq:characteristic_example}
\tau = \frac{1}{2}\frac{\ell^{2}}{D},
\end{gather} 
which is attained at the right end point ($x = \ell$) (Fig.~\ref{fig:mat_representations}). This problem demonstrates that MAT produces a characteristic timescale for the single-layer (homogeneous) diffusion model described by Eqs.~(\ref{eq:pde}) and (\ref{eq:ic_bcs}) that is proportional to the usual ratio $\ell^{2}/D$ \cite{crank_1975}. This result provides us with the necessary motivation to apply the concept of MAT to multilayer diffusion processes.

\section{Characteristic timescales for multilayer diffusion}
\label{sec:multilayer}
We now consider a model of diffusion in a heterogeneous medium partitioned into $m$ layers, such that $0 = x_{0} < x_{1} < \hdots < x_{m-1} < x_{m} = \ell = \sum_{i=1}^{m}\ell_{i}$, where $[x_{i-1},x_{i}]$ and $\ell_{i}$ defines layer $i$ and its width, respectively. The governing equations are
\begin{gather}
\label{eq:multdiff_pde}
R_{i}\frac{\partial u_{i}}{\partial t} = D_{i}\frac{\partial^{2}u_{i}}
{\partial x^{2}},\quad x_{i-1} < x < x_{i},\quad t > 0,\\
\label{eq:multdiff_ic}
u_{i} = 0,\quad t = 0,\\ 
\label{eq:multdiff_bcL}
u_{1} = u_{b},\quad x = x_{0},\\
\label{eq:multdiff_bcR}
\frac{\partial u_{m}}{\partial x} = 0,\quad x = x_{m},
\end{gather}
where $i = 1,\hdots,m$. When modelling mass diffusion, $D_{i}$ is the diffusivity in layer $i$ and $R_{i} = 1$ in each layer. For heat conduction, $R_{i}$ and $D_{i}$ are the volumetric heat capacity and thermal conductivity in layer $i$, respectively. In groundwater flow models, $R_{i}$ is the storage coefficient and $D_{i}$ is the hydraulic conductivity in layer $i$. 

At the interfaces between adjacent layers, we study the following types of internal boundary conditions which are commonly applied at the interfaces \cite{carr_2016a,kaoui_2018,hickson_2009_part1,liu_1998}:
\begin{enumerate}[(a)]
\item Type A
\begin{gather}
\label{eq:multdiff_typeA1}
D_{i}\frac{\partial u_{i}}{\partial x} = H_{i}(\theta_{i}u_{i+1} - u_{i}),\quad\text{$x = x_{i}$},\\
\label{eq:multdiff_typeA2}
D_{i+1}\frac{\partial u_{i+1}}{\partial x} = H_{i}(\theta_{i}u_{i+1} - u_{i}),\quad\text{$x = x_{i}$},
\end{gather}
\item Type B
\begin{gather}
\label{eq:multdiff_typeB1}
u_{i} = \theta_{i}u_{i+1},\quad\text{$x = x_{i}$},\\
\label{eq:multdiff_typeB2}
D_{i}\frac{\partial u_{i}}{\partial x} = D_{i+1}\frac{\partial u_{i+1}}{\partial x},\quad\text{$x = x_{i}$},
\end{gather}
\end{enumerate}
where $\theta_{i}>0$ and $H_{i}>0$ are, respectively, the partition and transfer coefficients at the interface between layers $i$ and $i+1$. \revision{For finite $H_{i}$, Eqs.~(\ref{eq:multdiff_typeA1})--(\ref{eq:multdiff_typeA2}) produce a discontinuity in the solution across the interface located at $x = x_{i}$. This behaviour finds application to several practical problems involving contact resistance across an interface, for example, modelling drug release from multi-layer capsules, where a thin coating layer is used to prevent fast release \cite{kaoui_2018}. On the other hand, Eqs.~(\ref{eq:multdiff_typeB1})--(\ref{eq:multdiff_typeB2}) maintain a constant ratio between the solution values across the interface and are useful in partitioned diffusion problems such as chemical transport in composite media \cite{trefry_1999}.} As Eqs.~(\ref{eq:multdiff_typeA1})--(\ref{eq:multdiff_typeA2}) are equivalent to Eqs.~(\ref{eq:multdiff_typeB1})--(\ref{eq:multdiff_typeB2}) in the case of infinite transfer coefficient ($H_{i}\rightarrow\infty$) or zero resistance at the interface ($H_{i}^{-1} \rightarrow 0$) \cite{carr_2017c,march_2017,gudnason_2018}, we consider only Type A conditions in the analysis presented in this section. 

We now apply the concept of MAT to the multilayer diffusion model (\ref{eq:multdiff_pde})--(\ref{eq:multdiff_typeA2}). Previously, \citet{hickson_thesis_2010} calculated MAT for a similar multilayer problem by first solving for the transient solution ($u_{i}(t;x)$ for $i = 1,\hdots,m$) using an orthogonal eigenfunction expansion and inserting a truncated form of the expansion into the multilayer analogue of Eq.~(\ref{eq:mat2}). The novelty of our analysis is that we show that the transient solution is not required to compute the multilayer analogues of Eq.~(\ref{eq:mat2}) and the characteristic timescale (\ref{eq:characteristic_timescale}) and this leads to simple formulas for the characteristic timescale as we will see in the next section. As with the single-layer problem of Section \ref{sec:mean_action_time}, we do however require the steady state solution of (\ref{eq:multdiff_pde})--(\ref{eq:multdiff_typeA2}), which is given by:
\begin{align}
\label{eq:multdiff_steady_state}
u_{i,\infty}(x) = \frac{u_{b}}{\prod_{k=1}^{i-1}\theta_{k}}.
\end{align}

For the multilayer problem, we define the following cumulative distribution functions:
\begin{equation}
\label{eq:multdiff_cdf}
F_{i}(t; x) := 1 - \frac{u_{i}(t;x)-u_{i,\infty}(x)}{u_{i,0}(x)-u_{i,\infty}(x)},
\end{equation}
for $i = 1,\hdots,m$. As such, MAT is piecewise continuous and defined in the $i$th layer as follows:
\begin{equation}
\label{eq:multdiff_mat1}
M_{i}(x) := \int_{0}^{\infty} tf_{i}(t;x)\,\mathrm{d}t,
\end{equation}
where:
\begin{align}
\label{eq:multdiff_pdf}
f_{i}(t; x) &:= \frac{\partial F_{i}(t;x)}{\partial t}.
\end{align}
In a similar manner to the development of Eq.~(\ref{eq:mat2}), combining (\ref{eq:multdiff_cdf})--(\ref{eq:multdiff_pdf}), applying integration by parts and noting that $\lim_{t\rightarrow\infty} t\left[u_{i}(t;x) - u_{i,\infty}(x)\right] = 0$ yields:
\begin{gather*}
M_{i}(x) = \int_{0}^{\infty} \frac{u_{i,\infty}(x)-u_{i}(t;x)}{u_{i,\infty}(x)-u_{0}(x)}\,\mathrm{d}t.
\end{gather*}
Following (\ref{eq:characteristic_timescale}), we define the characteristic timescale for a multilayer diffusion process by the maximum value of MAT over all layers:
\begin{gather}
\label{eq:multdiff_characteristic_timescale}
\tau = \max_{i=1,\hdots,m}\,\max_{x\in[x_{i-1},x_{i}]} M_{i}(x).
\end{gather}

The process of computing MAT in each layer without requiring the transient solution is now described. Firstly, define $\overline{M}_{i}(x) := M_{i}(x)h_{i}(x)$:
\begin{gather}
\label{eq:multdiff_mbar}
\overline{M}_{i}(x) = \int_{0}^{\infty} u_{i,\infty}(x)-u_{i}(t;x)\,\mathrm{d}t,
\end{gather}
where $h_{i}(x) := u_{i,\infty}(x)-u_{0}(x)$. Differentiating (\ref{eq:multdiff_mbar}) twice with respect to $x$ and using (\ref{eq:multdiff_pde}) yields:
\begin{gather}
\label{eq:mat_ode}
\overline{M}_{i}''(x) = -\frac{R_{i}}{D_{i}}\int_{0}^{\infty}\frac{\partial u_{i}}{\partial t}\,\mathrm{d}t.
\end{gather}
Performing the integration and inserting the initial condition (\ref{eq:multdiff_ic}) and steady state solution (\ref{eq:multdiff_steady_state}) yields the multilayer analogue of the differential equation (\ref{eq:mat_de}) satisfied by MAT:
\begin{gather}
\label{eq:mat_ode}
\overline{M}_{i}''(x) = -\frac{R_{i}}{D_{i}}\frac{u_{b}}{\prod_{k=1}^{i-1}\theta_{i}},\quad x_{i-1} < x < x_{i}.
\end{gather}
The general solution of this differential equation is given by:
\begin{gather}
\label{eq:Mbar}
\overline{M}_{i}(x) = \alpha_{i} + \beta_{i}x -\frac{R_{i}u_{b}}{2D_{i}\prod_{k=1}^{i-1}\theta_{k}}x^{2},
\end{gather}
where $\alpha_{i}$ and $\beta_{i}$ are integration constants that satisfy the linear system formulated by substituting the functions $\overline{M}_{1}(x),\hdots,\overline{M}_{m}(x)$ (\ref{eq:Mbar}) into the boundary conditions:
\begin{gather}
\label{eq:mat_bc1}
\overline{M}_{1}(x_{0}) = 0,\\
\label{eq:mat_ic1}
H_{i}^{-1}D_{i}\overline{M}_{i}'(x_{i}) = \theta_{i}\overline{M}_{i+1}(x_{i}) - \overline{M}_{i}(x_{i}),\\
H_{i}^{-1}D_{i+1}\overline{M}_{i+1}'(x_{i}) = \theta_{i}\overline{M}_{i+1}(x_{i}) - \overline{M}_{i}(x_{i}),\\
\label{eq:mat_bc2}
\overline{M}_{m}'(x_{m}) = 0,
\end{gather}
We remark that Eqs.~(\ref{eq:mat_bc1})--(\ref{eq:mat_bc2}) are derived using Eq.~(\ref{eq:multdiff_mbar}) and the boundary and interface conditions (\ref{eq:multdiff_bcL})--(\ref{eq:multdiff_typeA2}). For example, Eq.~(\ref{eq:mat_ic1}) is derived by differentiating (\ref{eq:multdiff_mbar}) with respect to $x$ and considering the expression:
\begin{align}
H_{i}^{-1}D_{i}\overline{M}_{i}'(x_{i}) = H_{i}^{-1}\int_{0}^{\infty} \Bigl[D_{i}u_{i,\infty}'(x_{i})-D_{i}\frac{\partial u_{i}}{\partial x}(x_{i},t)\Bigr]\,\mathrm{d}t.
\end{align}
Inserting the interface conditions (\ref{eq:multdiff_typeA1})--(\ref{eq:multdiff_typeA2}) (including the corresponding interface conditions satisfied by $u_{i,\infty}(x)$) yields the stated result in Eq.~(\ref{eq:mat_ic1}).

Once $\alpha_{i}$ and $\beta_{i}$ ($i= 1,\hdots,m$) are computed by solving Eqs.~(\ref{eq:mat_bc1})--(\ref{eq:mat_bc2}), the function (\ref{eq:Mbar}) is identified in every layer and MAT in the $i$th layer is calculated as $M_{i}(x) = \overline{M}_{i}(x)/h_{i}(x)$. Note that for the external boundary conditions (\ref{eq:multdiff_bcL})--(\ref{eq:multdiff_bcR}), the maximum (\ref{eq:multdiff_characteristic_timescale}) is attained at the right end point ($x = x_{m}$), giving a characteristic timescale of $\tau = M_{m}(x_{m})$. 

\section{Results}
\label{sec:results}
To illustrate our approach for calculating the characteristic timescale, we first consider the multilayer diffusion model (\ref{eq:multdiff_pde})--(\ref{eq:multdiff_typeA2}) with $m = 2$, $\theta_{1} = 1$ and $H_{1}\rightarrow\infty$. In this case, Eqs.~(\ref{eq:mat_bc1})--(\ref{eq:mat_bc2}) can be solved symbolically in a computer algebra system such as Maple \cite{maple16}, which leads to the following functions:
\begin{align}
\label{eq:matx1}
M_{1}(x) &= -\frac{1}{2}\frac{x^{2}}{D_{1}} + \frac{\left(\ell_{1}+\ell_{2}\right)x}{D_{1}},\\ 
\label{eq:matx2}
M_{2}(x) &= -\frac{1}{2}\frac{x^{2}}{D_{2}} + \frac{\left(\ell_{1}+\ell_{2}\right)x}{D_{2}} - \frac{1}{2}\frac{\ell_{1}(\ell_{1}+2\ell_{2})\left(D_{1}-D_{2}\right)}{D_{1}D_{2}},
\end{align}
which describe how MAT varies with position $x$ within the first and second layers, respectively. Both $M_{1}(x)$ and $M_{2}(x)$ are monotonically increasing on $(x_0,x_1)$ and $(x_1,x_2)$, which makes intuitive sense, as one would expect MAT to be greater the further away from the left end point ($x = x_{0}$) where the disturbance is introduced into the system via the boundary condition (\ref{eq:multdiff_bcL}). As previously mentioned, for the external boundary conditions (\ref{eq:multdiff_bcL})--(\ref{eq:multdiff_bcR}), the maximum value of MAT across both layers occurs at the right end point ($x = x_{m}$). Evaluating (\ref{eq:matx2}) at $x = x_{2} = \ell_{1}+\ell_{2}$ produces the following formula for the characteristic timescale of the process:
\begin{align}
\label{eq:mat_example}
\tau = \frac{1}{2}\frac{\ell_{1}^{2}}{D_{1}}+\frac{1}{2}\frac{\ell_{2}^{2}}{D_{2}} + \frac{\ell_{1}\ell_{2}}{D_{1}}.
\end{align}
This simple expression describes how the diffusive timescale varies with the thicknesses and diffusivities of the first and second layers. Note that in either of the cases (i) $D_{1} = D_{2} = D$ (ii) $\ell_{1} = 0$, $D_{2} = D$ or (iii) $\ell_{2} = 0$, $D_{1} = D$, Eq.~(\ref{eq:mat_example}) simplifies to the homogeneous (single-layer) result (\ref{eq:characteristic_example}). An important observation is that the order of the two layers is important when characterising the timescale as a different value of $\tau$ is obtained if $D_{1}$ and $D_{2}$ are interchanged in Eq.~(\ref{eq:mat_example}). Moreover, the weight of $1/D_{1}$ ($\ell_{1}^{2}/2+\ell_{1}\ell_{2}$) is greater than the weight of $1/D_{2}$ ($\ell_{2}^{2}/2$) provided $\ell_{1}/\ell_{2} > (\sqrt{2}-1) \approx 0.4142$. This means that the value of $D_{1}$ has a much greater influence on the characteristic timescale for this process: even if the first layer is half the width of the second layer ($\ell_{1}/\ell_{2} = 0.5$) the weighting of $1/D_{1}$ is greater than the weight of $1/D_{2}$.

\begin{table*}[t]
\centering
\text{$m=2$ layers}\\
\begin{tabular}{|c|l|}
\hline
Interface condition & Characteristic timescale\\
\hline
\rule{0pt}{0.8cm}Type A & \quad$\displaystyle\tau=\frac{1}{2}\frac{R_{1}\ell_{1}^{2}}{D_{1}}+\frac{1}{2}\frac{R_{2}\ell_{2}^{2}}{D_{2}} + \frac{R_{2}\ell_{1}\ell_{2}}{D_{1}\theta_{1}} + \revision{\frac{R_{2}\ell_{2}}{H_{1}\theta_{1}}}$\quad\\[0.4cm]
Type B & \quad$\displaystyle \tau=\frac{1}{2}\frac{R_{1}\ell_{1}^{2}}{D_{1}}+\frac{1}{2}\frac{R_{2}\ell_{2}^{2}}{D_{2}} + \frac{R_{2}\ell_{1}\ell_{2}}{D_{1}\theta_{1}}$\quad\\[0.4cm]
\hline

\multicolumn{2}{c}{\text{$m=3$ layers}}\\
\hline
Interface condition & Characteristic timescale\\
\hline
\rule{0pt}{0.8cm}Type A & \quad$\displaystyle\tau = \frac{1}{2}\frac{R_{1}\ell_{1}^{2}}{D_{1}}+\frac{1}{2}\frac{R_{2}\ell_{2}^{2}}{D_{2}}+\frac{1}{2}\frac{R_{3}\ell_{3}^{2}}{D_{3}} + \frac{R_{2}\ell_{1}\ell_{2}}{D_{1}\theta_{1}} + \frac{R_{3}\ell_{2}\ell_{3}}{D_{2}\theta_{2}} + \frac{R_{3}\ell_{1}\ell_{3}}{D_{1}\theta_{1}\theta_{2}} +  \revision{\frac{R_{2}\ell_{2}}{H_{1}\theta_{1}}$ \\[0.3cm] & \hspace{1.2cm}$\displaystyle+ \frac{R_{3}\ell_{3}}{H_{2}\theta_{2}} + \frac{R_{3}\ell_{3}}{H_{1}\theta_{1}\theta_{2}}}$\\
\rule{0pt}{0.8cm}Type B & \quad$\displaystyle\tau = \frac{1}{2}\frac{R_{1}\ell_{1}^{2}}{D_{1}}+\frac{1}{2}\frac{R_{2}\ell_{2}^{2}}{D_{2}}+\frac{1}{2}\frac{R_{3}\ell_{3}^{2}}{D_{3}} + \frac{R_{2}\ell_{1}\ell_{2}}{D_{1}\theta_{1}} + \frac{R_{3}\ell_{2}\ell_{3}}{D_{2}\theta_{2}} + \frac{R_{3}\ell_{1}\ell_{3}}{D_{1}\theta_{1}\theta_{2}}$\\[0.4cm]
\hline

\multicolumn{2}{c}{\text{$m$ layers}}\\
\hline
Interface condition & Characteristic timescale\\
\hline
\revision{\rule{0pt}{0.8cm}Type A} & \quad\revision{$\displaystyle\tau = \frac{1}{2}\sum_{i=1}^{m}\frac{R_{i}\ell_{i}^{2}}{D_{i}} + \sum_{i=1}^{m-1}\sum_{j=i+1}^{m}\frac{R_{j}\ell_{j}}{\prod_{k=i}^{j-1}\theta_{k}}\left[\frac{\ell_{i}}{D_{i}} + \frac{1}{H_{i}}\right]$}\\[0.5cm]
\rule{0pt}{0.8cm}Type B & \quad$\displaystyle\tau = \frac{1}{2}\sum_{i=1}^{m}\frac{R_{i}\ell_{i}^{2}}{D_{i}} + \sum_{i=1}^{m-1}\sum_{j=i+1}^{m}\frac{R_{j}\ell_{i}\ell_{j}}{D_{i}\prod_{k=i}^{j-1}\theta_{k}}$\\[0.5cm]
\hline
\end{tabular}
\caption{\textbf{Characteristic timescale formulas.} Characteristic timescales for the multilayer diffusion model (\ref{eq:multdiff_pde})--(\ref{eq:multdiff_bcR}) with Type A (\ref{eq:multdiff_typeA1})--(\ref{eq:multdiff_typeA2}) or Type B (\ref{eq:multdiff_typeB1})--(\ref{eq:multdiff_typeB2}) interface conditions.}
\label{tab:mat_2layers}
\end{table*}

Repeating the above working, Table \ref{tab:mat_2layers} summarises the characteristic timescales of the multilayer diffusion model (\ref{eq:multdiff_pde})--(\ref{eq:multdiff_typeA2}) for two-layer ($m=2$) and three-layer ($m=3$) diffusion processes. These results provide simple formulas that explain how modifying the internal boundary condition at the interface affects the characteristic timescale (\ref{eq:multdiff_characteristic_timescale}). For the two-layer case, since $H_{1} > 0$, the characteristic timescale is always larger for Type A interface conditions (\ref{eq:multdiff_typeA1})--(\ref{eq:multdiff_typeA2}) than Type B interface conditions (\ref{eq:multdiff_typeB1})--(\ref{eq:multdiff_typeB2}) with equality obtained in the limit as $H_{i}\rightarrow\infty$. Moreover, the characteristic timescale increases for decreasing values of the contact transfer coefficient $H_{1}$ and partition coefficient $\theta_{1}$. Both of these observations make intuitive sense as the time to reach steady state is slowed for decreasing values of $H_{1}$ and $\theta_{1}$. The storage coefficients, $R_{1}$ and $R_{2}$, also have a retardation effect with the characteristic timescale decreasing for decreasing values of $R_{1}$ and $R_{2}$ in all cases. Similar observations can be drawn for the three-layer case.

Generalisation of the two-layer and three-layer characteristic timescales to an arbitrary number of layers is also presented in Table \ref{tab:mat_2layers} for \revision{both types of interface conditions}. These expressions demonstrate that the timescale is characterised by the sum of the single-layer timescale (\ref{eq:characteristic_example}) across each layer ($\ell_{i}^{2}/(2D_{i})$ for $i = 1,\hdots,m$), modified appropriately to incorporate the storage coefficients, with a correction term accounting for the pairwise coupling that exists between each layer and every other layer. \revision{Finally, using the general $m$-layer result for Type B interface conditions (Table \ref{tab:mat_2layers}) with $R_{i} = 1$ ($i = 1,\hdots,m$) and $\theta_{k} = 1$ ($k = 1,\hdots,m-1$) generalises, to an arbitrary number of layers, the characteristic timescale given in Eq.~(\ref{eq:mat_example}) for pure diffusion in a two-layer medium with perfect contact at the interfaces:
\begin{align}
\tau = \frac{1}{2}\sum_{i=1}^{m}\frac{\ell_{i}^{2}}{D_{i}} + \sum_{i=1}^{m-1}\sum_{j=i+1}^{m}\frac{\ell_{i}\ell_{j}}{D_{i}}.
\end{align}
 } 

\begin{figure*}[!t]
\centering
\includegraphics[width=\textwidth]{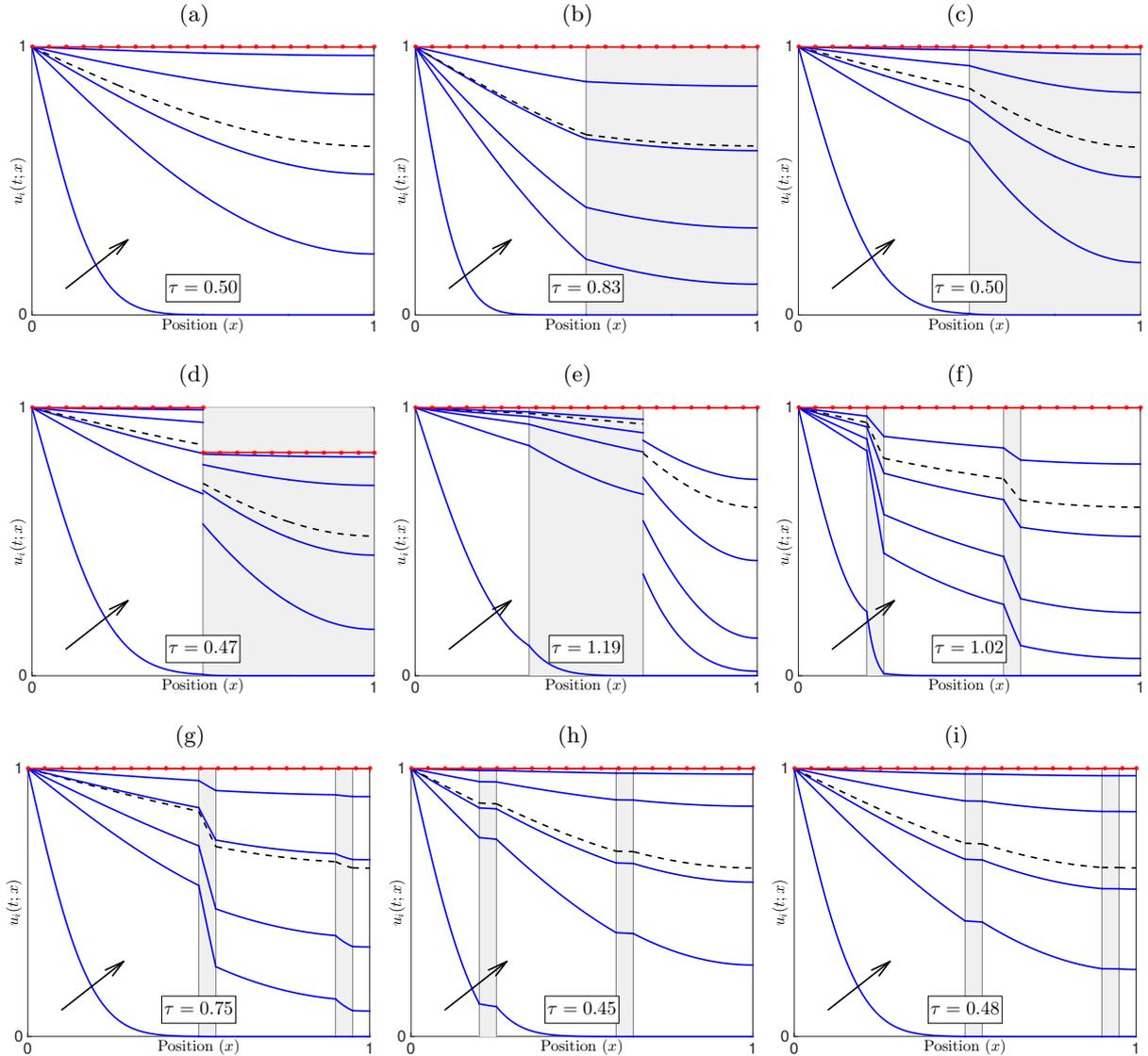}
\caption{\textbf{Characterising the time required for a multilayer diffusion process to reach steady state.}~Plot of the transient solution $u_{i}(t;x)$ at \revision{$t = [0.01,0.2,0.4,0.8,1.5]$ (blue solid lines) and $t = \tau$ (black dashed line)} for $i = 1,\hdots,m$ of the multilayer diffusion model (\ref{eq:multdiff_pde})--(\ref{eq:multdiff_typeA2}) for the nine different diffusive processes (A)--(I) discussed in Section \ref{sec:results}. Each plot includes the characteristic timescale ($\tau$), the steady state solution $u_{i,\infty}(x)$ (\ref{eq:multdiff_steady_state}) for $i = 1,\hdots,m$ \revision{(red dash-dot line)} and a black arrow indicating the direction of increasing time $t$.}
\label{fig:results}
\end{figure*}

We now demonstrate how our approach can be used to characterise and compare the timescales of different multilayer diffusive processes. Consider the multilayer diffusion model (\ref{eq:multdiff_pde})--(\ref{eq:multdiff_typeA2}) with $u_{b}=1$ and the following parameters:
\begin{enumerate}[(A)]
\item $m=1$, $D_{1} = 1$, $R_{1} = 1$, $\ell_{1} = 1$.
\item $m=2$, $[D_{1},D_{2}] = [0.5,1.5]$, $[R_{1},R_{2}] = [1,1]$, $H_{1}\rightarrow\infty$, $\theta_{1} = 1$, $[\ell_{1},\ell_{2}] = [0.5,0.5]$.
\item Same as (B) but with $[D_{1},D_{2}] = [1.5,0.5]$.
\item Same as (C) but with $\theta_{1} = 1.2$.
\item $m=3$, $[D_{1},D_{2},D_{3}] = [2,1,1]$, $[H_{1},H_{2}] \rightarrow [\infty,1]$, $[\ell_{1},\ell_{2},\ell_{3}] = [1/3,1/3,1/3]$. $R_{i} = 1$ and $\theta_{i} = 1$ for all $i$.
\item $m=5$, $[D_{1},D_{2},D_{3},D_{4},D_{5}] = [1,0.1,1,0.1,1]$, $[\ell_{1},\ell_{2},\ell_{3},\ell_{4},\ell_{5}] = [0.2,0.05,0.35,0.05,0.35]$.\\ $R_{i} = 1$, $\theta_{i} = 1$ and $H_{i}\rightarrow\infty$ for all $i$.
\item Same as (F) but with $[\ell_{1},\ell_{2},\ell_{3},\ell_{4},\ell_{5}] = [0.5,0.05,0.35,0.05,0.05]$.
\item Same as (F) but with $[D_{1},D_{2},D_{3},D_{4},D_{5}] = [1,10,1,10,1]$.
\item Same as (G) but with $[D_{1},D_{2},D_{3},D_{4},D_{5}] = [1,10,1,10,1]$.
\end{enumerate}
A practical question of interest is which of these processes takes the longest amount of time to reach steady state and which takes the shortest amount of time. To answer such questions, we evaluate the characteristic timescales (Table \ref{tab:mat_2layers}), which take on the following values (rounded to two decimal places) for the diffusion processes listed above:
\begin{alignat*}{3}
&\text{(A)}\hspace{0.2cm}\tau = 0.50 &\qquad& \text{(B)}\hspace{0.2cm}\tau = 0.83 &\qquad& \text{(C)}\hspace{0.2cm} \tau = 0.50\\
&\text{(D)}\hspace{0.2cm}\tau = 0.47 &\qquad& \text{(E)}\hspace{0.2cm}\tau = \revision{1.19} &\qquad& \text{(F)}\hspace{0.2cm} \tau = 1.02\\
&\text{(G)}\hspace{0.2cm}\tau = 0.75 &\qquad& \text{(H)}\hspace{0.2cm}\tau = 0.45 &\qquad& \text{(I)}\hspace{0.2cm} \tau = 0.48.
\end{alignat*}
To assess the validity of using the characteristic timescale as an indicator of the time required to reach steady state, in Fig.~\ref{fig:results} we plot the transient solution $u_{i}(t;x)$ ($i = 1,\hdots,m$) of the multilayer diffusion model (\ref{eq:multdiff_pde})--(\ref{eq:multdiff_typeA2}), \revision{calculated using the semi-analytical approach given by \citet{carr_2017c}}, versus position $x$ for all nine diffusion processes. Moving from $D_{1}=1$ (A) to $D_{1}=0.5$ and $D_{2}=1.5$ (B) produces a larger characteristic timescale, which is consistent with (B) taking longer than (A) to reach steady state  (Fig.~\ref{fig:results}). On the other hand, process (C), which simply reverses the order of the layers of (B), gives rise to an identical characteristic timescale to process (A). This is consistent with the observation from Fig.~\ref{fig:results} that processes (A) and (C) take the same amount of time to effectively reach steady state. Process (D) is the same as (C) with the exception that the partition coefficient is modified from 1.0 to 1.2. This gives rise to a discontinuity in the solution at the interface between the two layers due to Eq.~(\ref{eq:multdiff_typeB1}), a modified steady state solution and a transient solution that narrowly attains its steady state quicker as is reflected in the magnitude of the characteristic timescale (0.47 for (D) compared to 0.50 for (C)). Process (E) involves a finite transfer coefficient at the interface between the second and third layers ($x = 2/3$), as evident in the discontinuous nature of the solution across the interface. This case also clearly takes the longest to reach steady state and this is correctly captured by the characteristic timescale as it is largest for (E) out of all nine diffusion processes. Processes (F) and (G) consider diffusion through a medium consisting of five layers with two low diffusive thin layers. In (F), the thin layers are taken as the intervals $x = [0.2,0.25]$ and $x = [0.6,0.65]$ while (G) shifts these layers by a length of 0.3 in the positive $x$ direction. The location of the thin low diffusive layers has a huge impact on the rate of transfer across the medium with the transfer rate of (F) noticeably slower than (G) and this is correctly captured by the characteristic timescale (1.02 for (F) compared to 0.75 for (G)). Processes (H) and (I) include high diffusive rather than low diffusive thin layers, which clearly accelerates the transition to steady state compared to processes (F) and (G). However, the location of these thin high diffusive layers has almost negligible affect on the transition time with process (H) only marginally faster than (I). This behaviour is again correctly captured by the characteristic timescales, with the value for (H) slightly less than (I) (0.45 versus 0.48, respectively). Finally, we note that ordering processes (A)--(I) from smallest to largest characteristic timescale is equivalent to ordering them in terms of the time taken to visually reach steady state.

\section{Conclusions}
\label{sec:conclusions}
This paper provides a simple tool for characterising the timescale of multilayer diffusion processes. Our approach utilises the concept of mean action time (MAT), which relies on identifying a suitable cumulative distribution function representing the transition of the diffusive process from initial to steady state. Notably, by extending ideas for single-layer (homogeneous) diffusion, we showed how MAT can be calculated without requiring the transient solution of the underlying multilayer diffusion model. The characteristic timescale is then defined as the maximum value of MAT over the medium. For different choices of internal and external boundary conditions, our approach produces simple formulas for characterising the timescale of multilayer diffusion processes that explain how the timescale is influenced by model parameters such as the diffusivities, storage coefficients and lengths of the individual layers, and the partition and transfer coefficients that apply at the interfaces between adjacent layers. Finally, we demonstrated how these formulas provide useful insight when assessing how the model parameters affect how long it takes for a multilayer diffusion process to reach steady state.

Throughout this paper, we have considered only the case of a Dirichlet boundary condition at the left end point ($x = x_0$) and a zero Neumann boundary condition at the right end point ($x = x_{m}$), however, our approach easily extends to other types of boundary conditions. For example, suppose the Dirichlet boundary condition at the left end point (\ref{eq:multdiff_bcL}) is replaced with the following Newton-type condition:
\begin{gather}
\label{eq:robin_bc}
D_{1}\frac{\partial u_{1}}{\partial x} = \sigma(u_{1} - u_{b}),\quad x = x_{0}.
\end{gather}
where $\sigma > 0$. In this case, repeating the analysis presented for the two-layer problem ($m=2$) at the start of Section \ref{sec:results} yields a characteristic timescale of
\begin{gather}
\label{eq:robin_bc_timescale}
\tau = \frac{1}{2}\frac{\ell_{1}^{2}}{D_{1}}+\frac{1}{2}\frac{\ell_{2}^{2}}{D_{2}} + \frac{\ell_{1}\ell_{2}}{D_{1}}+\frac{(\ell_{1}+\ell_{2})}{\sigma}.
\end{gather}
Eq.~(\ref{eq:robin_bc_timescale}) increases with decreasing $\sigma$, which is consistent with the boundary condition (\ref{eq:robin_bc}) as smaller values of $\sigma$ result in a slower transition from initial to steady state. Moreover, in the limit $\sigma\rightarrow\infty$, where the boundary conditions (\ref{eq:robin_bc}) and (\ref{eq:multdiff_bcL}) are equivalent, the characteristic timescales (\ref{eq:robin_bc_timescale}) and (\ref{eq:mat_example}) are also equivalent.

While we have considered only multilayer diffusion in Cartesian coordinates, we note that our methodology carries over to other linear transport processes (advection-diffusion, reaction-diffusion) and other coordinate systems (cylindrical and spherical). The applicability to processes in spherical coordinates, in particular, should prove useful in assessing drug delivery systems involving multilayer spherical capsules \cite{kaoui_2018}.

\revision{Finally, we note that the method, based on the higher-moments of the distribution, for calculating the time required for the transient solution to transition to within a small specified tolerance of the steady-state solution (as presented in the author's previous work \cite{carr_2017b} for single-layer diffusion) can be extended to the multilayer case. However, the use of higher moments for layered problems involves more tedious algebra and leads to complicated expressions. For this reason, in this paper, we have considered only the first moment (MAT), which provides simple elegant expressions for characterising the timescale that are of greater practical use.}

\section*{Acknowledgements}
This research was funded by the Australian Research Council (DE150101137).

\bibliographystyle{plainnat}
\bibliography{references}

\end{document}